\documentclass[twocolumn]{aastex63}
\providecommand{\bjdtdb}{\ensuremath{\rm {BJD_{TDB}}}}

\providecommand{\msun}{\ensuremath{\,M_\Sun}}
\providecommand{\rsun}{\ensuremath{\,R_\Sun}}
\providecommand{\lsun}{\ensuremath{\,L_\Sun}}

\providecommand{\me}{\ensuremath{\,M_\oplus}}
\providecommand{\re}{\ensuremath{\,R_\oplus}}
\def\gtaprx{ \mathrel{ \vcenter{
      \offinterlineskip \hbox{$>$}
      \kern 0.3ex \hbox{$\sim$}    } } }

\def\ltaprx{ \mathrel{ \vcenter{
      \offinterlineskip \hbox{$<$}
      \kern 0.3ex \hbox{$\sim$}    } } }

\providecommand{\fave}{\langle F \rangle}
\providecommand{\fluxcgs}{10$^9$ erg s$^{-1}$ cm$^{-2}$}

\shorttitle{Transits of Known Planets Orbiting a Naked-Eye Star}
\shortauthors{Stephen R. Kane et al.}

\begin{document}

\title{Transits of Known Planets Orbiting a Naked-Eye Star}


\author[0000-0002-7084-0529]{Stephen R. Kane}
\affiliation{Department of Earth and Planetary Sciences, University of
  California, Riverside, CA 92521, USA}
\email{skane@ucr.edu}

\author[0000-0002-5224-247X]{Sel\c{c}uk Yal\c{c}{\i}nkaya}
\affiliation{Ankara University, Graduate School of Natural and Applied
  Sciences, Department of Astronomy and Space Sciences, Tandogan,
  TR-06100, Turkey}

\author[0000-0002-4047-4724]{Hugh P. Osborn}
\affiliation{Aix-Marseille Université, CNRS, CNES, Laboratoire
  d'Astrophysique de Marseille, France}
\affiliation{Department of Physics and Kavli Institute for
  Astrophysics and Space Research, Massachusetts Institute of
  Technology, Cambridge, MA 02139, USA}
\affiliation{NCCR/PlanetS, Centre for Space \& Habitability,
  University of Bern, Bern, Switzerland}

\author[0000-0002-4297-5506]{Paul A. Dalba}
\altaffiliation{NSF Astronomy and Astrophysics Postdoctoral Fellow}
\affiliation{Department of Earth and Planetary Sciences, University of
  California, Riverside, CA 92521, USA}

\author[0000-0002-0091-9362]{Louise D. Nielsen}
\affiliation{Observatoire de Gen\`eve, Universit\'e de Gen\`eve, 51
  ch. des Maillettes, 1290 Sauverny, Switzerland}

\author[0000-0001-7246-5438]{Andrew Vanderburg}
\altaffiliation{NASA Sagan Fellow}
\affiliation{Department of Astronomy, The University of Texas at
  Austin, Austin, TX 78712, USA}

\author[0000-0003-4603-556X]{Teo Mo\v{c}nik}
\affiliation{Department of Earth and Planetary Sciences, University of
  California, Riverside, CA 92521, USA}

\author[0000-0003-0595-5132]{Natalie R. Hinkel}
\affiliation{Southwest Research Institute, 6220 Culebra Rd, San
  Antonio, TX 78238, USA}

\author[0000-0001-7968-0309]{Colby Ostberg}
\affiliation{Department of Earth and Planetary Sciences, University of
  California, Riverside, CA 92521, USA}

\author[0000-0002-6191-459X]{Ekrem Murat Esmer}
\affiliation{Ankara University, Faculty of Science, Department of
  Astronomy and Space Sciences, Tandogan, TR-06100, Turkey}

\author[0000-0001-7576-6236]{St\'ephane Udry}
\affiliation{Observatoire de Gen\`eve, Universit\'e de Gen\`eve, 51
  ch. des Maillettes, 1290 Sauverny, Switzerland}

\author[0000-0002-3551-279X]{Tara Fetherolf}
\affiliation{Department of Earth and Planetary Sciences, University of
  California, Riverside, CA 92521, USA}

\author[0000-0002-4746-0181]{\"Ozg\"ur Ba\c{s}t\"urk}
\affiliation{Ankara University, Faculty of Science, Department of
  Astronomy and Space Sciences, Tandogan, TR-06100, Turkey}


\author[0000-0003-2058-6662]{George R. Ricker}
\affiliation{Department of Physics and Kavli Institute for
  Astrophysics and Space Research, Massachusetts Institute of
  Technology, Cambridge, MA 02139, USA}

\author[0000-0001-6763-6562]{Roland Vanderspek}
\affiliation{Department of Physics and Kavli Institute for
  Astrophysics and Space Research, Massachusetts Institute of
  Technology, Cambridge, MA 02139, USA}

\author[0000-0001-9911-7388]{David W. Latham}
\affiliation{Center for Astrophysics ${\rm \mid}$ Harvard {\rm \&}
  Smithsonian, 60 Garden Street, Cambridge, MA 02138, USA}

\author[0000-0002-6892-6948]{Sara Seager}
\affiliation{Department of Physics and Kavli Institute for
  Astrophysics and Space Research, Massachusetts Institute of
  Technology, Cambridge, MA 02139, USA}
\affiliation{Department of Earth, Atmospheric and Planetary Sciences,
  Massachusetts Institute of Technology, Cambridge, MA 02139, USA}
\affiliation{Department of Aeronautics and Astronautics, MIT, 77
  Massachusetts Avenue, Cambridge, MA 02139, USA}

\author[0000-0002-4265-047X]{Joshua N. Winn}
\affiliation{Department of Astrophysical Sciences, Princeton
  University, 4 Ivy Lane, Princeton, NJ 08544, USA}

\author[0000-0002-4715-9460]{Jon M. Jenkins}
\affiliation{NASA Ames Research Center, Moffett Field, CA, 94035, USA}


\author[0000-0002-1199-9759]{Romain Allart}
\affiliation{Observatoire de Gen\`eve, Universit\'e de Gen\`eve, 51
  ch. des Maillettes, 1290 Sauverny, Switzerland}

\author[0000-0002-5726-7000]{Jeremy Bailey}
\affiliation{School of Physics, University of New South Wales, Sydney,
  NSW 2052, Australia}

\author[0000-0003-4733-6532]{Jacob L. Bean}
\affil{Department of Astronomy \& Astrophysics, University of Chicago,
  5640 South Ellis Avenue, Chicago, IL 60637, USA}

\author[0000-0002-7613-393X]{Francois Bouchy}
\affiliation{Observatoire de Gen\`eve, Universit\'e de Gen\`eve, 51 ch. des Maillettes, 1290 Sauverny, Switzerland}

\author[0000-0003-1305-3761]{R. Paul Butler}
\affiliation{Earth \& Planets Laboratory, Carnegie Institution of
  Washington, NW, Washington, DC, 20015-1305, USA}

\author[0000-0002-4588-5389]{Tiago L. Campante}
\affiliation{Instituto de Astrof\'{\i}sica e Ci\^{e}ncias do
  Espa\c{c}o, Universidade do Porto, Rua das Estrelas, 4150-762 Porto,
  Portugal}
\affiliation{Departamento de F\'{\i}sica e Astronomia, Faculdade de
  Ci\^{e}ncias da Universidade do Porto, Rua do Campo Alegre, s/n,
  4169-007 Porto, Portugal}

\author[0000-0003-0035-8769]{Brad D. Carter}
\affiliation{Centre for Astrophysics, University of Southern
  Queensland, Toowoomba, QLD 4350, Australia}

\author[0000-0002-6939-9211]{Tansu Daylan}
\altaffiliation{Kavli Fellow}
\affiliation{Department of Physics and Kavli Institute for
  Astrophysics and Space Research, Massachusetts Institute of
  Technology, Cambridge, MA 02139, USA}

\author[0000-0001-6036-0225]{Magali Deleuil}
\affiliation{Aix-Marseille Université, CNRS, CNES, Laboratoire
  d'Astrophysique de Marseille, France}

\author[0000-0001-9289-5160]{Rodrigo F. Diaz}
\affiliation{International Center for Advanced Studies (ICAS) and
  ICIFI (CONICET), ECyT-UNSAM, Campus Miguelete, 25 de Mayo y Francia
  (1650), Buenos Aires, Argentina}

\author[0000-0002-9332-2011]{Xavier Dumusque}
\affiliation{Observatoire de Gen\`eve, Universit\'e de Gen\`eve, 51
  ch. des Maillettes, 1290 Sauverny, Switzerland}

\author[0000-0001-9704-5405]{David Ehrenreich}
\affiliation{Observatoire de Gen\`eve, Universit\'e de Gen\`eve, 51
  ch. des Maillettes, 1290 Sauverny, Switzerland}

\author[0000-0002-1160-7970]{Jonathan Horner}
\affiliation{Centre for Astrophysics, University of Southern
  Queensland, Toowoomba, QLD 4350, Australia}

\author[0000-0001-8638-0320]{Andrew W. Howard}
\affiliation{Department of Astronomy, California Institute of
  Technology, Pasadena, CA 91125, USA}

\author[0000-0002-0531-1073]{Howard Isaacson}
\affiliation{Department of Astronomy, University of California,
  Berkeley, CA 94720, USA}
\affiliation{Centre for Astrophysics, University of Southern
  Queensland, Toowoomba, QLD 4350, Australia}

\author[0000-0003-0433-3665]{Hugh R.A. Jones}
\affiliation{Centre for Astrophysics Research, University of
  Hertfordshire, Hatfield, Herts AL10 9AB, UK}

\author[0000-0002-2607-138X]{Martti H. Kristiansen}
\affiliation{Brorfelde Observatory, Observator Gyldenkernes Vej 7,
  DK-4340 T\o{}ll\o{}se, Denmark}
\affiliation{DTU Space, National Space Institute, Technical University
  of Denmark, Elektrovej 327, DK-2800 Lyngby, Denmark}

\author[0000-0001-7120-5837]{Christophe Lovis}
\affiliation{Observatoire de Gen\`eve, Universit\'e de Gen\`eve, 51
  ch. des Maillettes, 1290 Sauverny, Switzerland}

\author[0000-0002-2909-0113]{Geoffrey W. Marcy}
\affiliation{Department of Astronomy, University of California,
  Berkeley, CA 94720, USA}

\author[0000-0001-5630-1396]{Maxime Marmier}
\affiliation{Observatoire de Gen\`eve, Universit\'e de Gen\`eve, 51
  ch. des Maillettes, 1290 Sauverny, Switzerland}

\author[0000-0003-2839-8527]{Simon J. O'Toole}
\affiliation{Australian Astronomical Observatory, North Ryde, NSW
  2113, Australia}
\affiliation{Australian Astronomical Optics, Faculty of Science and
  Engineering, Macquarie University, North Ryde, NSW 2113, Australia}

\author{Francesco Pepe}
\affiliation{Observatoire de Gen\`eve, Universit\'e de Gen\`eve, 51
  ch. des Maillettes, 1290 Sauverny, Switzerland}

\author[0000-0003-1080-9770]{Darin Ragozzine}
\affil{Department of Physics and Astronomy, N283 ESC, Brigham Young
  University, Provo, UT 84602, USA}

\author{Damien S\'egransan}
\affiliation{Observatoire de Gen\`eve, Universit\'e de Gen\`eve, 51
  ch. des Maillettes, 1290 Sauverny, Switzerland}

\author[0000-0002-7595-0970]{C.G. Tinney}
\affiliation{Exoplanetary Science at USNW, School of Physics , UNSW
  Sydney, Sydney 2052, Australia}

\author[0000-0002-0569-1643]{Margaret C. Turnbull}
\affiliation{SETI Institute, Carl Sagan Center for the Study of Life
  in the Universe, Off-Site: 2613 Waunona Way, Madison, WI 53713, USA}

\author[0000-0001-9957-9304]{Robert A. Wittenmyer}
\affiliation{Centre for Astrophysics, University of Southern
  Queensland, Toowoomba, QLD 4350, Australia}

\author[0000-0001-7294-5386]{Duncan J. Wright}
\affiliation{Centre for Astrophysics, University of Southern
  Queensland, Toowoomba, QLD 4350, Australia}

\author[0000-0001-6160-5888]{Jason T. Wright}
\affiliation{Department of Astronomy \& Astrophysics and Center for
  Exoplanets and Habitable Worlds and Penn State Extraterrestrial
  Intelligence Center, 525 Davey Laboratory, The Pennsylvania State
  University, University Park, PA 16802, USA}


\begin{abstract}

Some of the most scientifically valuable transiting planets are those
that were already known from radial velocity (RV) surveys. This is
primarily because their orbits are well characterized and they
preferentially orbit bright stars that are the targets of RV
surveys. The Transiting Exoplanet Survey Satellite ({\it TESS})
provides an opportunity to survey most of the known exoplanet systems
in a systematic fashion to detect possible transits of their
planets. HD~136352 (Nu$^2$~Lupi) is a naked-eye ($V = 5.78$) G-type
main-sequence star that was discovered to host three planets with
orbital periods of 11.6, 27.6, and 108.1 days via RV monitoring with
the HARPS spectrograph. We present the detection and characterization
of transits for the two inner planets of the HD~136352 system,
revealing radii of $1.482^{+0.058}_{-0.056}$~$R_\oplus$ and
$2.608^{+0.078}_{-0.077}$~$R_\oplus$ for planets b and c,
respectively. We combine new HARPS observations with RV data from
Keck/HIRES and the AAT, along with {\it TESS} photometry from Sector
12, to perform a complete analysis of the system parameters. The
combined data analysis results in extracted bulk density values of
$\rho_b = 7.8^{+1.2}_{-1.1}$~gcm$^{-3}$ and $\rho_c =
3.50^{+0.41}_{-0.36}$~gcm$^{-3}$ for planets b and c, respectively,
thus placing them on either side of the radius valley. The combination
of the multi-transiting planet system, the bright host star, and the
diversity of planetary interiors and atmospheres means this will
likely become a cornerstone system for atmospheric and orbital
characterization of small worlds.

\end{abstract}



\section{Introduction}
\label{intro}

The discovery of transiting exoplanets has enabled a plethora of
science not accessible through other exoplanet detection
techniques. Transiting planets orbiting bright stars are especially
important in furthering our knowledge of planetary systems because
they offer unique windows to comparative exoplanetology. First, they
allow for a measurement of both the planetary mass and radius, and
thereby to place constraints on the planet interior structure. Second,
they are amenable to atmospheric characterization through transmission
spectroscopy \citep[e.g.,][]{sing2016,kempton2018}, to secondary
eclipse measurements \citep[e.g.,][]{kreidberg2019b}, and to orbital
geometry characterization through the Rossiter-McLaughlin effect
\citep[e.g.,][]{fabrycky2009}. In multi-planet systems, they also
allow for a deepened understanding of the system architecture through
planet-planet dynamics accessible through the modeling of transit
timing variations \citep[e.g.,][]{jontofhutter2016}.

The Transit Ephemeris Refinement and Monitoring Survey (TERMS) has
been operating since 2008 with the primary goal of detecting transits
for known radial velocity (RV) exoplanets \citep{kane2009c}. The
appeal of transits for known RV planets is that their orbits are
already characterized and their host stars are relatively
bright. Well-known examples of RV planets later found to transit
include HD~209458b \citep{charbonneau2000,henry2000a}, HD~189733b
\citep{bouchy2005c}, and HD~80606b
\citep{naef2001b,fossey2009,garciamelendo2009a,laughlin2009a}. Though
the TERMS survey successfully discovered new planets \citep{wang2012},
destroyed old planets \citep{kane2016a}, characterized numerous host
stars \citep[e.g.,][]{dragomir2012a,hinkel2015b}, and ruled out
transits \citep[e.g.,][]{kane2011b,kane2011c,pilyavsky2011,henry2013},
the primary science goal was largely impeded by ground-based
observational window functions \citep{vonbraun2009}. However, the
Transiting Exoplanet Survey Satellite ({\it TESS}) has observed most
of the sky during the primary mission \citep{ricker2015}, including
the known exoplanet hosts. The survey strategy of {\it TESS} thus
provides a space-based means to systematically examine all of the
known RV systems for potential transits of their planets that pass
through inferior conjunction during the {\it TESS} observing window
\citep{kane2008b,dalba2019c}. This has been demonstrated through the
transit detection of an additional planet in the pi~Mensae system
\citep{huang2018} and the known RV planet in the HD~118203 system
\citep{pepper2020}.

HD~136352 (also known as Nu$^2$~Lupi, LHS~395, GJ~582, and HIP~75181)
is a G-type main-sequence star that was observed for nearly 11 years
using the High Accuracy Radial velocity Planet Searcher (HARPS)
spectrograph \citep{pepe2000}. Analysis of the HARPS data by
\citet{udry2019} uncovered the signatures of three planets orbiting
the star with periods and minimum masses in the range of 11--110~days
and 5--10~$M_\oplus$ respectively. The host star was observed by {\it
  TESS} during Sector~12 of the primary mission. A {\it TESS} Guest
Investigator (GI) program designed to monitor the known hosts (PI:
Kane) immediately detected transits of the two inner planets (b and
c). As of 2020 June 9, the HD~136352 system is one of less than 100
naked-eye exoplanet host stars in the sky, according to data from the
NASA Exoplanet Archive \citep{akeson2013}. HD~136352 is now also one
of only three naked-eye stars that host more than one transiting
planet, the other two systems being HD~219134
\citep{motalebi2015,vogt2015,gillon2017b} and HR~858
\citep{vanderburg2019}. In all cases, the measurement accuracy
correlates with the stellar magnitude, and therefore the search for
transiting planets around bright stars is of paramount importance.

Here we present the detection of transits for the two inner planets of
the HD~136352 system from {\it TESS} photometry, and provide a
combined analysis of all available RV and photometric data. In
Section~\ref{system} we describe the transit probabilities of the
planets and the properties of the host star. Section~\ref{obs}
provides details regarding the RV observations of the star and the
detrending of the {\it TESS} photometry. A description of the data
analysis is provided in Section~\ref{analysis}, including a combined
fit for all available data and a discussion of the location of the
planets with respect to the overall demographics of exoplanets. In
Section~\ref{atmos} we present a discussion of the potential for
atmospheric characterization of the transiting planets. Suggestions
for future work and concluding remarks are provided in
Section~\ref{conclusions}.


\section{System Properties}
\label{system}

The HD~136352 system is known to host three planets with orbital
periods of 11.6, 27.6, and 108.1 days. The planetary orbits are
near-circular in nature, and the detailed properties of the new
parameters provided by this work may be found in
Section~\ref{analysis}. Since the planets have short orbital periods,
they also have relatively high geometric transit probabilities
\citep{kane2008b,stevens2013}. We calculate the a priori transit
probabilities as 4.88\%, 2.77\%, and 1.19\% for the b, c, and d
planets, respectively. A detailed analysis of transit probabilities for
known RV exoplanets by \citet{dalba2019c} resulted in a prediction of
$\sim$3 transit detections during the {\it TESS} primary mission. Our
reporting of transits for two of the HD~136352 planets, combined with
the detected transit for HD~118203b \citep{pepper2020}, brings the
total number of RV planets revealed to be transiting in line with the
\citet{dalba2019c} predictions. Note that the {\it TESS} extended
mission will likely lead to further transit detections of known RV
planets whose orbital periods are longer than {\it TESS} observations
of their host star during the primary mission.

The host star HD~136352 is a G3/5V star that has been
spectroscopically observed dozens of times over the last three decades
\citep{hinkel2014}. Compiling the observed stellar parameters, to be
used as priors in Section~\ref{exofast}, we found that the median
stellar radius $R_* = 1.02 \pm 0.02$~$R_{\odot}$, $T_\mathrm{eff} =
5692 \pm 218$~K, and $\log g = 4.39 \pm 0.33$, such that the
uncertainties reflect the \textit{spread} or range in all of the
measured values. The iron content, or [Fe/H], ranges from -0.16 dex
\citep{carretta2000a} to -0.46 dex \citep{francois1986a}, with a
median value of $-0.29 \pm 0.15$~dex, where all observations were
solar renormalized to \citet{lodders2009}. Based on the 32 other
elemental abundances reported in the Hypatia Catalog
\citep{hinkel2014}, it is clear that HD~136352 is a relatively
metal-poor star, consistent with the star's thick-disk
kinematics. Some of the elements, particularly those in the iron peak
and beyond the iron peak nucleosynthetic groups, have abundances that
are more dramatically sub-solar, such as [Cr/H] = $-0.31 \pm
0.14$~dex, [Mn/H] = $-0.48 \pm 0.25$~dex, and [Y/H] = $-0.34 \pm
0.07$~dex. On the other hand, the $\alpha$-elements are closer to
solar, such as [O/H] = $-0.02 \pm 0.15$~dex and [Mg/H] = $-0.04 \pm
0.15$~dex. The average of the $\alpha$-elements, particularly C, O,
Mg, Si, S, Ca, and Ti, is [$\alpha$/H] = -0.12 dex. The C/O molar
fraction for HD~136352 is 0.35, where stars with a C/O ratio
$\sim$0.8--1.0 are likely to produce geodynamically inactive planets
\citep{bond2010a,unterborn2014,hinkel2018}.


\section{Observations}
\label{obs}

The observational data of the system considered here include almost 20
years of precision RV measurements  and one sector of {\it TESS}
photometry during Cycle 1. The star is identified in the {\it TESS}
Input Catalog (TIC) as TIC~136916387 \citep{stassun2018c,stassun2019}.


\subsection{Radial Velocities}
\label{rvs}

\begin{figure*}
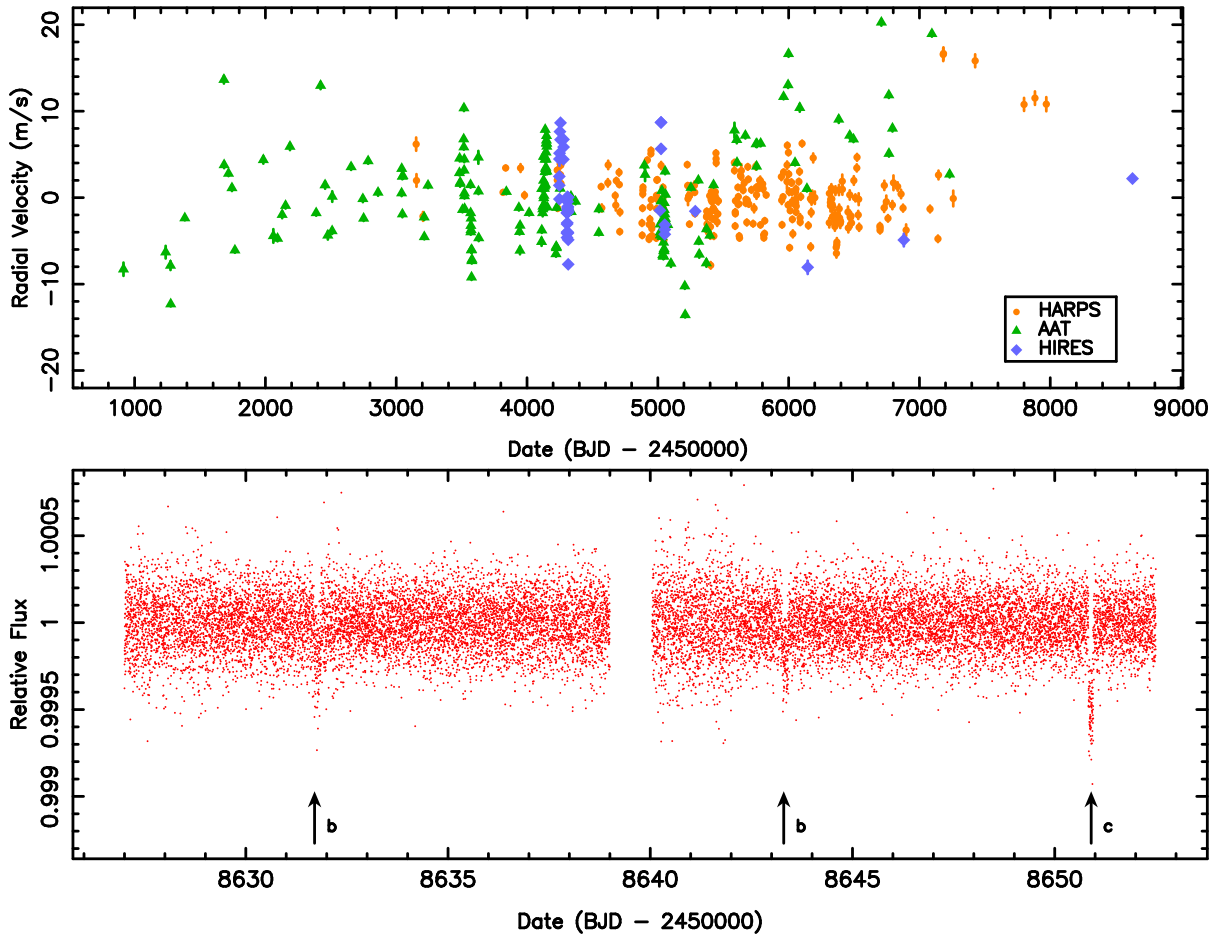

  \begin{center}
    \includegraphics[angle=270,width=16.0cm]{f01a.ps} \\
    \includegraphics[angle=270,width=16.0cm]{f01b.ps}
  \end{center}
  \caption{Primary data sources used in this analysis. {\it Top:} The
    combined RV data spanning a period of 21 years, acquired using the
    HARPS (orange circles), AAT/UCLES (green triangles), and
    Keck/HIRES (blue diamonds) instruments. {\it Bottom:} The Sector
    12 {\it TESS} photometry, with vertical arrows indicating the
    location of the two transits for planet b and the single transit
    for planet c.}
  \label{fig:data}
\end{figure*}

The RV data used for this analysis were acquired from three different
observing facilities. The first dataset consists of 246 RV
measurements obtained over a period of 13.2 years using the HARPS
spectrograph, of which 240 measurements were previously published by
\citet{udry2019} when announcing the discovery of the HD~136352
system. The full details of the instrument and observations may be
found in \citet{udry2019} and references therein. Note that the most
recent (6) HARPS measurements were acquired after an instrument
upgrade and so were treated as an independent dataset in the combined
fit to the data described in Section~\ref{exofast}. The second dataset
consists of 169 RV measurements obtained over a period of 17.3 years
using the UCLES high-resolution spectrograph \citep{diego1990} on the
3.9m Anglo-Australian Telescope (AAT). The instrument uses an iodine
absorption cell to provide wavelength calibration from 5000 to
6200\,\AA, by embedding iodine absorption lines on the stellar
spectrum \citep{valenti1995b,butler1996}. The AAT RV observations were
conducted as part of the Anglo-Australian Planet Search, described in
more detail by \citet{wittenmyer2020b} and references therein. The
third dataset consists of 43 RV measurements obtained over a period of
12.0 years using the HIRES echelle spectrograph on the Keck I
telescope \citep{vogt1994}, of which 23 measurements were previously
published by \citet{howard2016}. The combined RV dataset are shown in
the top panel of Figure~\ref{fig:data} and a subset of 10 RVs from
each instrument are provided in Table~\ref{vels}. The mean measurement
uncertainties are 0.42, 1.27, and 1.17 m/s for the HARPS, AAT, and
HIRES data sets respectively. Although the highest of these mean
uncertainties is associated with the AAT data, the AAT dataset also
has the longest time baseline, making it a valuable addition to the
analysis.

\begin{deluxetable}{lccc}
  \tablewidth{0pc}
  \tablecaption{\label{vels} HD~136352 radial velocities.}
  \tablehead{
    \colhead{Instrument} &
    \colhead{Date} &
    \colhead{RV} &
    \colhead{$\sigma$} \\
    \colhead{} &
    \colhead{(BJD -- 2450000)} &
    \colhead{(m/s)} &
    \colhead{(m/s)}
  }
  \startdata
HARPS & 3152.7661 &   6.188 &  0.770 \\
HARPS & 3154.6851 &   1.984 &  0.730 \\
HARPS & 3204.5378 &  -2.172 &  0.520 \\
HARPS & 3816.8345 &   0.609 &  0.280 \\
HARPS & 3836.7974 &   3.438 &  0.270 \\
HARPS & 3950.5459 &   3.422 &  0.460 \\
HARPS & 3980.4880 &   0.281 &  0.380 \\
HARPS & 4230.8179 &   3.180 &  0.300 \\
HARPS & 4231.7510 &   2.023 &  0.360 \\
HARPS & 4234.7314 &  -1.242 &  0.300 \\
AAT &  915.1653 &  -8.280 &  1.800 \\
AAT & 1237.2321 &  -6.310 &  3.380 \\
AAT & 1274.2877 &  -7.850 &  2.470 \\
AAT & 1276.1555 & -12.310 &  3.210 \\
AAT & 1384.0170 &  -2.360 &  1.840 \\
AAT & 1683.0382 &  13.650 &  2.100 \\
AAT & 1684.1084 &   3.790 &  2.000 \\
AAT & 1718.0880 &   2.790 &  2.100 \\
AAT & 1743.9812 &   1.120 &  1.840 \\
AAT & 1766.8840 &  -6.080 &  2.040 \\
HIRES & 6145.7617 &  -8.060 &  1.288 \\
HIRES & 6880.7524 &  -4.906 &  1.400 \\
HIRES & 5024.8408 &   5.643 &  1.286 \\
HIRES & 5024.8418 &   8.699 &  1.252 \\
HIRES & 5024.8428 &   8.706 &  1.286 \\
HIRES & 5052.8169 &  -3.129 &  1.219 \\
HIRES & 5052.8184 &  -4.245 &  1.199 \\
HIRES & 5052.8198 &  -3.695 &  1.236 \\
HIRES & 4246.9380 &   1.427 &  1.239 \\
HIRES & 4246.9390 &   2.457 &  1.268 \\
  \enddata
\tablenotetext{}{The full data set is available online.}
\end{deluxetable}


\subsection{{\it TESS} Photometry}
\label{phot}

The {\it TESS} spacecraft observed HD~136352 during Sector 12 of its
primary mission between May 21, 2019 and June 18, 2019. Because
HD~136352 is a bright, nearby dwarf star, images from pixels
surrounding the star were saved and downloaded every two minutes,
compared to 30 minute sampling for most of the sky. These images were
downlinked from the spacecraft, processed by the Science Processing
Operations Center (SPOC) pipeline (based at NASA Ames Research Center)
and searched for transits \citep{jenkins2016,jenkins2020a}. The SPOC
transiting planet search algorithm detected a possible transit-like
signal when it lined up one transit of HD~136352~b with the single
transit of HD~136352~c, but the signal was rejected by an automated
classification algorithm because the two transits have significantly
different depths. HD~136352 was therefore not alerted as a {\it TESS}
planet candidate host star. We subsequently identified the transits of
HD~136352~b and c in a visual inspection of the light curve, which
resulted in an allocated {\it TESS} Object of Interest (TOI) number of
2011.

The light curve of HD~136352 produced by the SPOC pipeline contains
residual systematic errors, so we extracted our own custom light curve
from the {\it TESS} pixel data. Our approach is very similar to the
one used by \citet{vanderburg2019} to produce a light curve of another
bright star, HR~858. We first extracted light curves of HD~136352 from
twenty different photometric apertures. We then removed instrumental
systematics by decorrelating each of the 20 light curves with other
time series via matrix inversion (while excluding points in-transit
from the fit). In particular, we decorrelated against the first and
second order time series of the means and standard deviations of the
engineering quaternion measurements within each exposure. We also
decorrelated against the high-frequency (band 3) common mode
systematics in the cotrending basis vectors calculated by the SPOC
Pre-search Data Conditioning module (PDC) and the time series of
background flux measurements
\citep{smith2012d,stumpe2012,stumpe2014}. In total, we we fit a model
with 46 free parameters to the 16,865 out-of-transit datapoints.
Finally, we calculated the point-to-point photometric scatter for each
of the 20 light curves and chose the one with the highest
precision. This procedure is described in more detail in Section 2.1
of \citet{vanderburg2019}.

The resulting decorrelated light curve still showed low-frequency
variability (likely a combination of both slow instrumental drifts and
astrophysical variability). We modeled these low-frequency trends with
a basis spline and simultaneously determined the spline function along
with the transit model parameters. We introduced discontinuities to
the basis spline at the times of spacecraft momentum dumps. We removed
the variability by dividing the best-fit spline from the light
curve. Note that the amplitude of the low-frequency variability is
less than a few hundred ppm with timescales greater than 1 day. Though
the variability amplitude is a large fraction of the transit depths
described in Section~\ref{analysis}, the timescales of the transits
are much shorter, and we were thus able to effectively remove the
variability with negligible effect on the subsequent data
analysis. The light curve resulting from our detrending, with both
instrumental systematics and stellar variability removed, is shown in
Figure~\ref{fig:data}.


\section{Data Analysis}
\label{analysis}

Here, we describe the extracted properties of the star and planets, as
well as their location within the context of the known exoplanet
population.


\subsection{Extraction of System Parameters}
\label{exofast}

\begin{figure}
  \begin{center}
      \includegraphics[width=8.5cm]{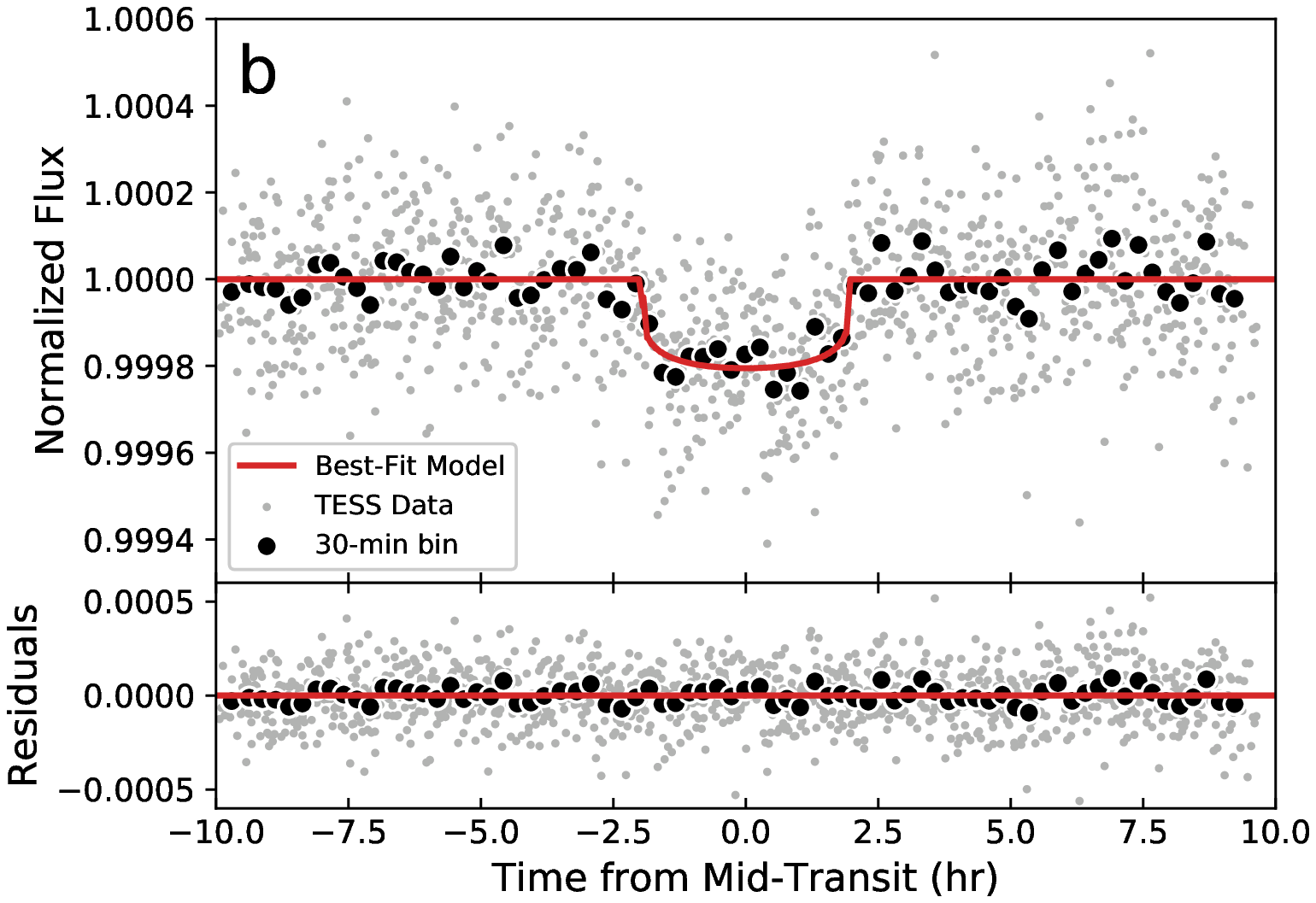} \\
      \includegraphics[width=8.5cm]{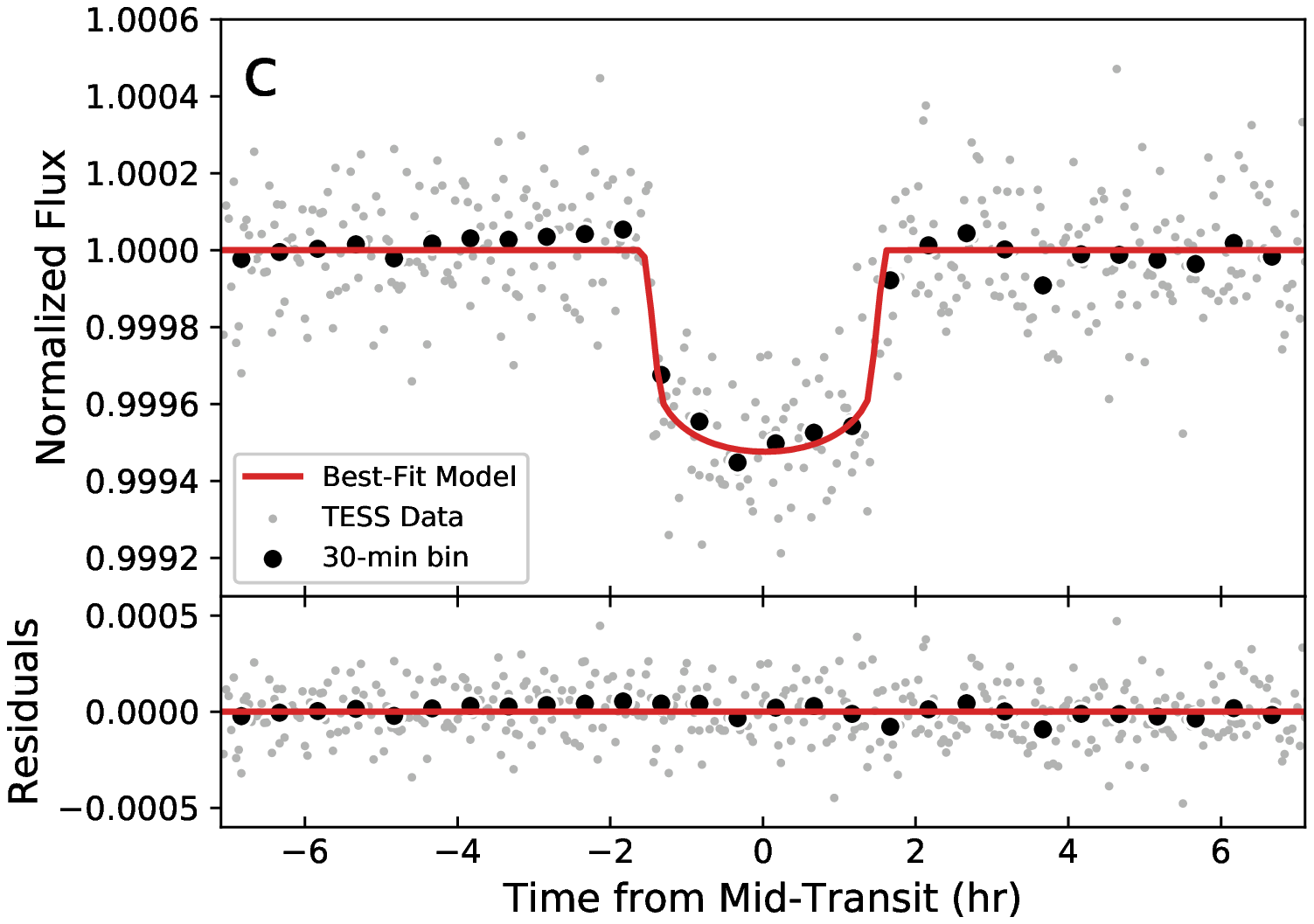} \\
  \end{center}
  \caption{Transit fits to the {\it TESS} photometry resulting from
    the {\sc EXOFAST} analysis, described in
    Section~\ref{exofast}. {\it Top:} Transit of the b planet where
    both transits have been folded on the orbital period of the
    planet. {\it Bottom:} Single transit of the c planet.}
  \label{fig:transits}
\end{figure}

\begin{figure}
  \begin{center}
      \includegraphics[width=8.5cm]{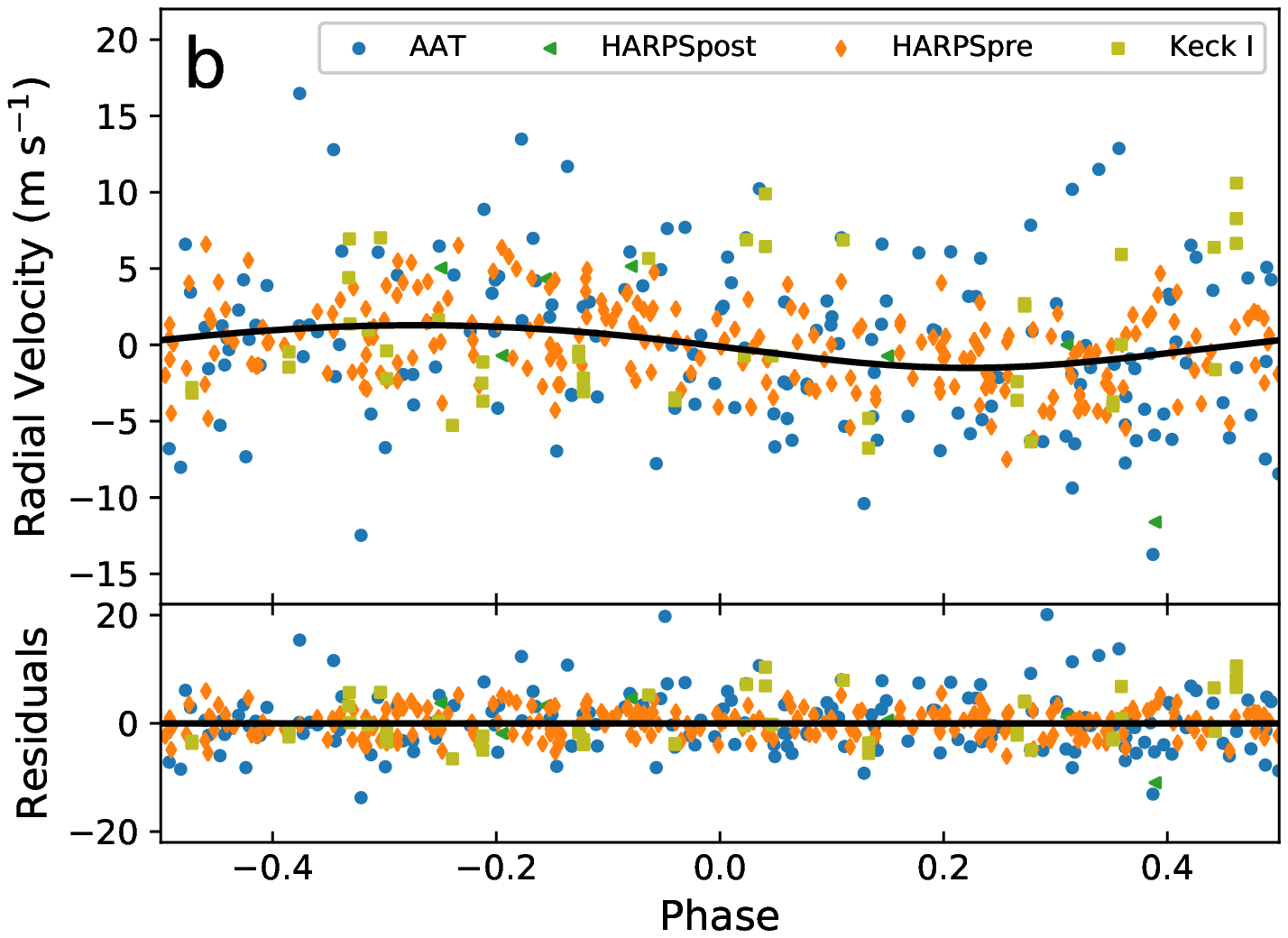} \\
      \includegraphics[width=8.5cm]{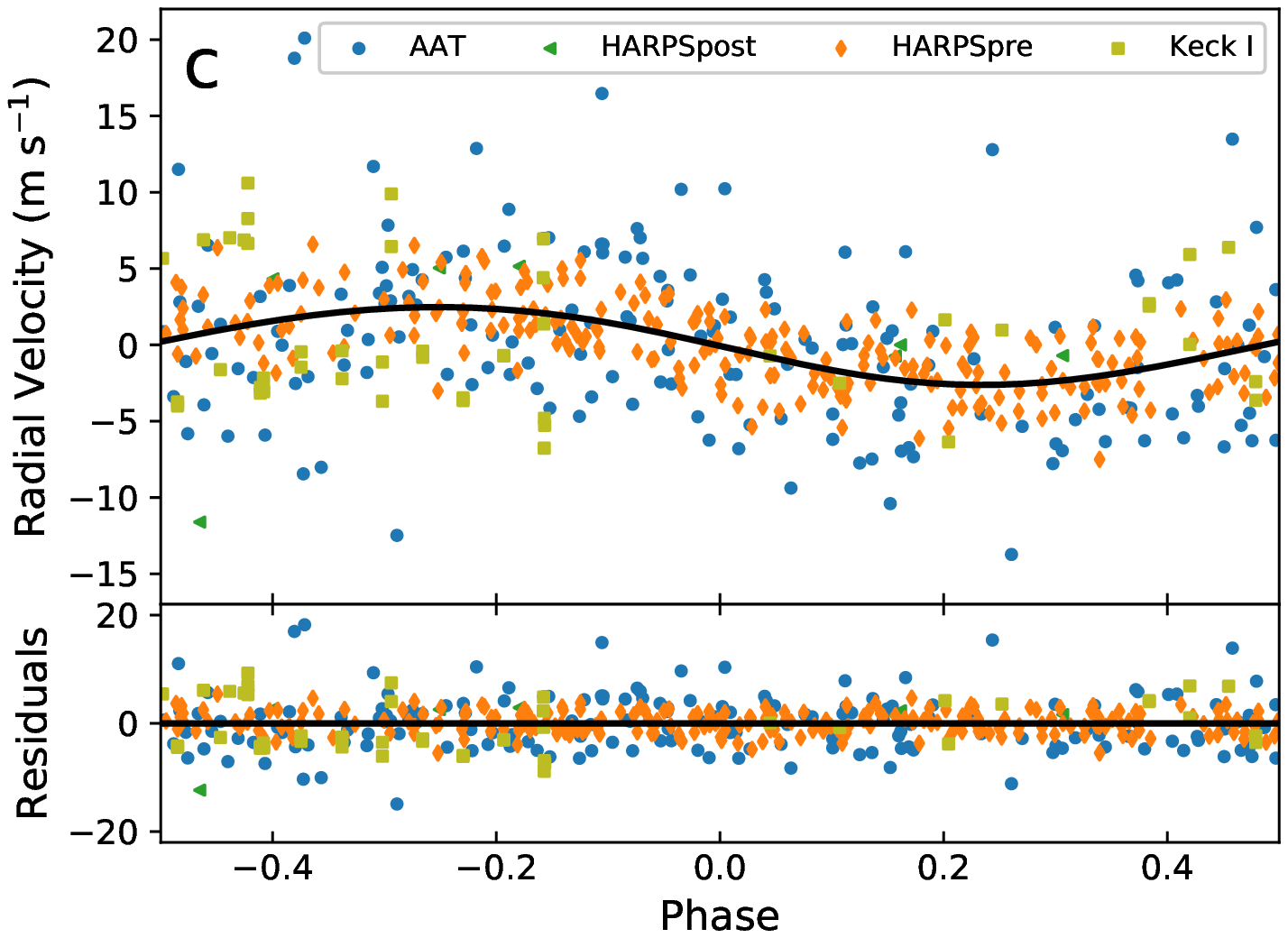} \\
      \includegraphics[width=8.5cm]{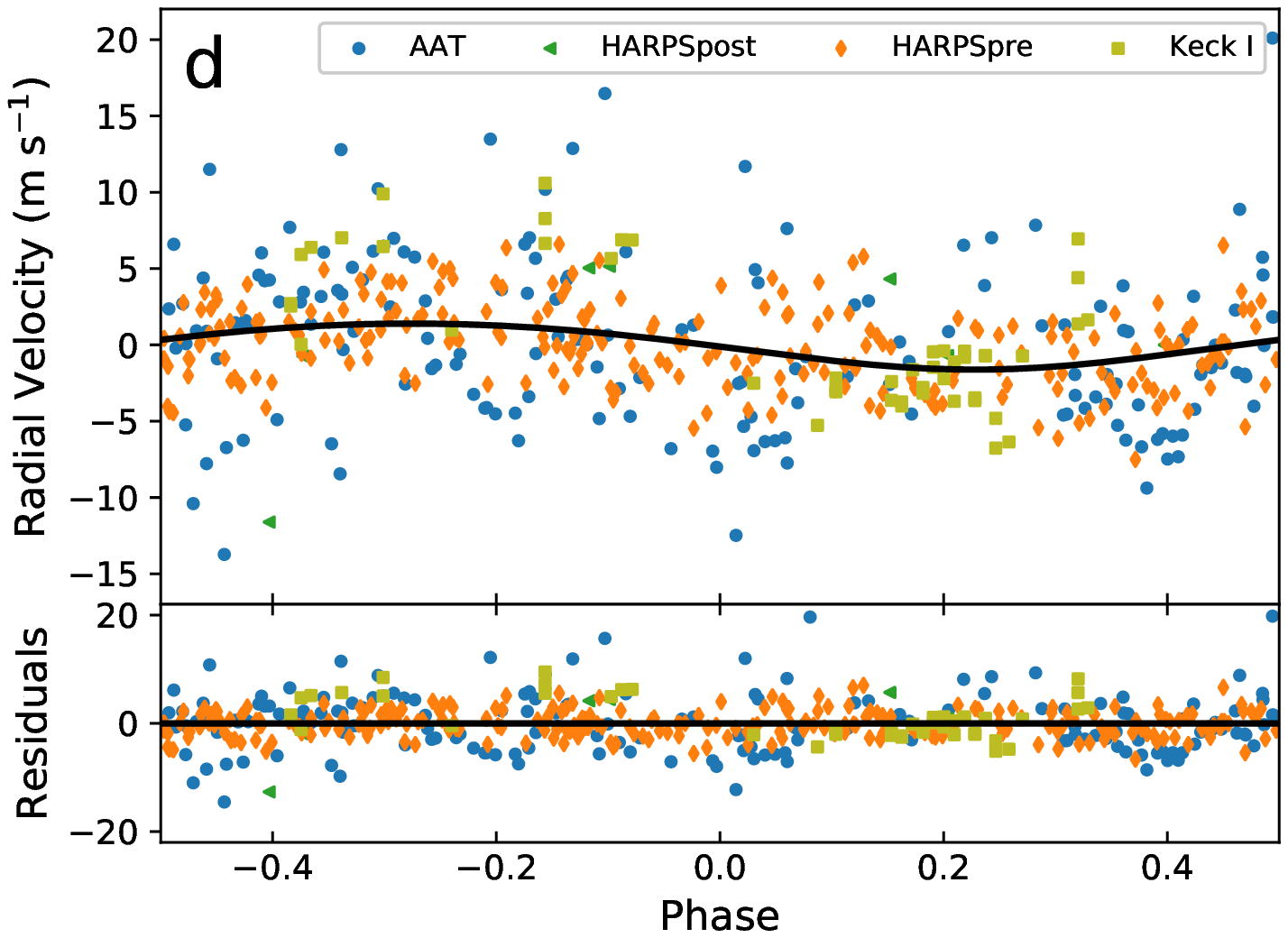} \\
  \end{center}
  \caption{Combined RV dataset after applying the {\sc EXOFAST} fits
    described in Section~\ref{exofast}. The RV data are folded on the
    orbital period for each planet, including the b (top), c (middle),
    and d (bottom) planets.}
  \label{fig:rvs}
\end{figure}

Using the data described in Section~\ref{obs}, we performed our
analysis using the {\sc EXOFASTv2}
tool\footnote{\url{https://github.com/jdeast/EXOFASTv2}}, described in
detail by \citet{eastman2013,eastman2020}. Following previous
applications of {\sc EXOFASTv2} \citep[e.g.,][]{dalba2020a}, we
conducted two fits to extract the system parameters. In the first, we
fit archival photometry of HD~136352 to modeled spectral energy
distributions (SED). We applied normal priors on the parameters
$R_{\star}$, $M_{\star}$, $T_{\rm eff}$, and [Fe/H] using the values
provided in Section~\ref{system}. We also included a normal prior on
parallax ($68.164\pm0.097$~mas) based on measurements from the second
data release of the {\it Gaia} mission \citep{brown2018} corrected for
the systematic offset discovered by \citet{stassun2018b}. Lastly, we
included an upper limit on the maximum line-of-sight extinction from
the reddening maps from \citet{schlafly2011}. This initial, SED-only
fit converged upon the following stellar parameters:
$R_{\star}=1.010\pm0.018 \; R_{\sun}$, $M_{\star}=0.923\pm0.077 \;
M_{\sun}$, $T_{\rm eff}=5851\pm110$~K, and [Fe/H]
$=-0.25\pm0.14$~dex. These parameters were then used as priors for a
global fit to the transit and RV data. We assessed convergence using
the default {\sc EXOFASTv2} statistics of $T_z$ \citep{ford2006d}, the
number of independent draws of the underlying posterior probability
distribution (convergence for $T_z>1000$ for each parameter), and GR,
the well-known Gelman-Rubin statistic \citep{gelman1992}, where
convergence is achieved for GR$<1.01$ for each parameter. The derived
stellar parameters from this global fit are provided in
Table~\ref{tab:star}. The planetary parameters are provided in
Table~\ref{tab:planet}. The zeroth-order, RV offsets found by the fit
were $-68709.038\pm0.094$, $-68697.2\pm3.0$, $0.17\pm0.38$, and
$-0.65\pm0.64$ m/s for the pre-upgrade HARPS, post-upgrade HARPS, AAT,
and HIRES data sets, respectively. The best-fit transit and RV models
are shown in Figure \ref{fig:transits} and Figure \ref{fig:rvs}.

Although transits of planets b and c were detected in the {\it TESS}
photometry, we found no evidence of a transit for planet d. The
question then is: do the photometric data rule out a transit for
planet d or did the planet not pass through inferior conjunction
during the {\it TESS} observing window? Using the derived orbital
properties of planet d from Table~\ref{tab:planet}, we found that the
nearest predicted inferior conjunction time occurs $\sim$21~days prior
to the commencement of Sector 12 observations. Thus, planet d may yet
be found to also transit the host star with further observations. The
inclinations of planets b and c are such that a perfectly coplanar
planet d would not be transiting, but even a tiny mutual inclination
of $\sim$0.3$^\circ$ would allow for a transit. In fact, planet d
would only be transiting for inclinations between 90$^\circ$ and
89.4$^\circ$, including grazing configurations, while assuming planet
d has a radius similar to planet c. Even so, conditioned on the fact
that b and c transit, the probability that d transits is about 20\%
for typical mutual inclinations of $\sim$1$^\circ$.

\begin{deluxetable*}{lcc}
\tablecaption{\label{tab:star} HD~136352 derived stellar parameters.}
\tablehead{\colhead{~~~Parameter} & \colhead{Units} & \colhead{Values}}
\startdata
~~~~$M_*$\dotfill &Mass (\msun)\dotfill &$0.906^{+0.055}_{-0.047}$\\
~~~~$R_*$\dotfill &Radius (\rsun)\dotfill &$1.012\pm0.018$\\
~~~~$L_*$\dotfill &Luminosity (\lsun)\dotfill &$1.081^{+0.088}_{-0.082}$\\
~~~~$\rho_*$\dotfill &Density (cgs)\dotfill &$1.234^{+0.098}_{-0.086}$\\
~~~~$\log{g}$\dotfill &Surface gravity (cgs)\dotfill &$4.385^{+0.029}_{-0.027}$\\
~~~~$T_{\rm eff}$\dotfill &Effective Temperature (K)\dotfill &$5850\pm100$\\
~~~~$[{\rm Fe/H}]$\dotfill & Proxy for metallicity (dex)\dotfill &$-0.25\pm0.12$\\
~~~~$[{\rm Fe/H}]_{0}$\dotfill & Proxy for initial metallicity \dotfill &$-0.18\pm0.11$\\
~~~~$Age$\dotfill &Age (Gyr)\dotfill &$8.2^{+3.2}_{-3.1}$\\
~~~~$A_V$\dotfill &V-band extinction (mag)\dotfill &$0.060^{+0.071}_{-0.043}$\\
~~~~$\sigma_{SED}$\dotfill &SED photometry error scaling \dotfill &$4.3^{+1.9}_{-1.1}$\\
~~~~$\varpi$\dotfill &Parallax (mas)\dotfill &$68.159\pm0.098$\\
~~~~$d$\dotfill &Distance (pc)\dotfill &$14.672\pm0.021$
\smallskip\\\multicolumn{2}{l}{Wavelength Parameters:}&TESS\smallskip\\
~~~~$u_{1}$\dotfill &linear limb-darkening coeff \dotfill &$0.275^{+0.039}_{-0.038}$\\
~~~~$u_{2}$\dotfill &quadratic limb-darkening coeff \dotfill &$0.285\pm0.035$\\
\enddata
\end{deluxetable*}

\begin{deluxetable*}{lcccc}
\tablecaption{\label{tab:planet} HD~136352 planetary parameters.}
\tablehead{\colhead{~~~Parameter} & \colhead{Units} & \multicolumn{3}{c}{Values}}
\startdata
\multicolumn{2}{l}{}&b&c&d\\
~~~~$P$\dotfill &Period (days)\dotfill &$11.57779^{+0.00091}_{-0.0011}$&$27.5909^{+0.0028}_{-0.0031}$&$107.63^{+0.18}_{-0.19}$\\
~~~~$R_p$\dotfill &Radius (\re)\dotfill &$1.482^{+0.058}_{-0.056}$&$2.608^{+0.078}_{-0.077}$&--\\
~~~~$M_p$\dotfill &Mass (\me)\dotfill &$4.62^{+0.45}_{-0.44}$&$11.29^{+0.73}_{-0.69}$&--\\
~~~~$T_C$\dotfill &Time of conjunction (\bjdtdb)\dotfill &$2458631.7672^{+0.0023}_{-0.0022}$&$2458650.8947^{+0.0011}_{-0.0010}$&$2458593.7^{+5.6}_{-5.5}$\\
~~~~$T_0$\dotfill &Optimal conjunction Time (\bjdtdb)\dotfill &$2458631.7672^{+0.0023}_{-0.0022}$&$2458650.8947^{+0.0011}_{-0.0010}$&$2455902.7^{+3.4}_{-2.7}$\\
~~~~$a$\dotfill &Semi-major axis (AU)\dotfill &$0.0969^{+0.0019}_{-0.0017}$&$0.1729^{+0.0034}_{-0.0030}$&$0.4285^{+0.0085}_{-0.0076}$\\
~~~~$i$\dotfill &Inclination (Degrees)\dotfill &$88.86^{+0.54}_{-0.30}$&$88.658^{+0.055}_{-0.057}$&--\\
~~~~$e$\dotfill &Eccentricity \dotfill &$0.079^{+0.068}_{-0.053}$&$0.037^{+0.039}_{-0.026}$&$0.075^{+0.085}_{-0.053}$\\
~~~~$\omega_*$\dotfill &Argument of Periastron (Degrees)\dotfill &$172^{+63}_{-67}$&$142^{+86}_{-92}$&$-175^{+79}_{-87}$\\
~~~~$T_{eq}$\dotfill &Equilibrium temperature (K)\dotfill &$911\pm18$&$682^{+14}_{-13}$&$433.3^{+8.6}_{-8.5}$\\
~~~~$K$\dotfill &RV semi-amplitude (m/s)\dotfill &$1.40\pm0.13$&$2.55\pm0.13$&$1.51\pm0.14$\\
~~~~$R_p/R_*$\dotfill &Radius of planet in stellar radii \dotfill &$0.01343^{+0.00044}_{-0.00045}$&$0.02363\pm0.00052$&--\\
~~~~$a/R_*$\dotfill &Semi-major axis in stellar radii \dotfill &$20.60^{+0.53}_{-0.49}$&$36.76^{+0.95}_{-0.87}$&$91.1^{+2.3}_{-2.2}$\\
~~~~$\delta$\dotfill &Transit depth (fraction)\dotfill &$0.000180\pm0.000012$&$0.000558^{+0.000025}_{-0.000024}$&--\\
~~~~$Depth$\dotfill &Flux decrement at mid transit \dotfill &$0.000180\pm0.000012$&$0.000558^{+0.000025}_{-0.000024}$&--\\
~~~~$\tau$\dotfill &Ingress/egress transit duration (days)\dotfill &$0.00260^{+0.00043}_{-0.00033}$&$0.0108\pm0.0011$&--\\
~~~~$T_{14}$\dotfill &Total transit duration (days)\dotfill &$0.1640^{+0.0048}_{-0.0045}$&$0.1337^{+0.0024}_{-0.0022}$&--\\
~~~~$T_{FWHM}$\dotfill &FWHM transit duration (days)\dotfill &$0.1613^{+0.0049}_{-0.0046}$&$0.1229^{+0.0023}_{-0.0022}$&--\\
~~~~$b$\dotfill &Transit Impact parameter \dotfill &$0.41^{+0.12}_{-0.20}$&$0.854^{+0.013}_{-0.016}$&--\\
~~~~$b_S$\dotfill &Eclipse impact parameter \dotfill &$0.408^{+0.077}_{-0.18}$&$0.863^{+0.056}_{-0.040}$&--\\
~~~~$\tau_S$\dotfill &Ingress/egress eclipse duration (days)\dotfill &$0.00261^{+0.00021}_{-0.00020}$&$0.0113^{+0.0035}_{-0.0017}$&--\\
~~~~$T_{S,14}$\dotfill &Total eclipse duration (days)\dotfill &$0.166^{+0.018}_{-0.014}$&$0.1321^{+0.0097}_{-0.020}$&--\\
~~~~$T_{S,FWHM}$\dotfill &FWHM eclipse duration (days)\dotfill &$0.163^{+0.018}_{-0.014}$&$0.121^{+0.011}_{-0.024}$&--\\
~~~~$\rho_p$\dotfill &Density (cgs)\dotfill &$7.8^{+1.2}_{-1.1}$&$3.50^{+0.41}_{-0.36}$&--\\
~~~~$logg_p$\dotfill &Surface gravity \dotfill &$3.313^{+0.053}_{-0.054}$&$3.211^{+0.038}_{-0.037}$&--\\
~~~~$\Theta$\dotfill &Safronov Number \dotfill &$0.0234^{+0.0023}_{-0.0022}$&$0.0581^{+0.0035}_{-0.0034}$&--\\
~~~~$\fave$\dotfill &Incident Flux (\fluxcgs)\dotfill &$0.155^{+0.013}_{-0.012}$&$0.0490^{+0.0040}_{-0.0038}$&$0.00790^{+0.00066}_{-0.00062}$\\
~~~~$T_P$\dotfill &Time of Periastron (\bjdtdb)\dotfill &$2458622.4^{+2.1}_{-1.9}$&$2458626.9\pm6.6$&$2458510^{+24}_{-25}$\\
~~~~$T_S$\dotfill &Time of eclipse (\bjdtdb)\dotfill &$2458625.63^{+0.37}_{-0.57}$&$2458636.93^{+0.39}_{-0.67}$&$2458536.8^{+6.6}_{-7.2}$\\
~~~~$e\cos{\omega_*}$\dotfill & \dotfill &$-0.047^{+0.050}_{-0.077}$&$-0.010^{+0.022}_{-0.038}$&$-0.030^{+0.049}_{-0.10}$\\
~~~~$e\sin{\omega_*}$\dotfill & \dotfill &$0.003^{+0.056}_{-0.057}$&$0.004^{+0.040}_{-0.027}$&$-0.002^{+0.050}_{-0.058}$\\
~~~~$M_p\sin i$\dotfill &Minimum mass (\me)\dotfill &$4.62^{+0.45}_{-0.43}$&$11.28^{+0.73}_{-0.69}$&$10.5^{+1.1}_{-1.0}$\\
~~~~$M_p/M_*$\dotfill &Mass ratio \dotfill &$0.0000153\pm0.0000014$&$0.0000373\pm0.0000021$&$0.0000346^{+0.0000033}_{-0.0000032}$\\
~~~~$d/R_*$\dotfill &Separation at mid transit \dotfill &$20.4\pm1.3$&$36.5^{+1.5}_{-1.6}$&$90.9^{+5.5}_{-5.7}$\\
\enddata
\end{deluxetable*}


\subsection{Exoplanet Demographics}
\label{demo}

\begin{figure*}
    \centering
        \begin{tabular}{cc}
    \includegraphics[width=\columnwidth]{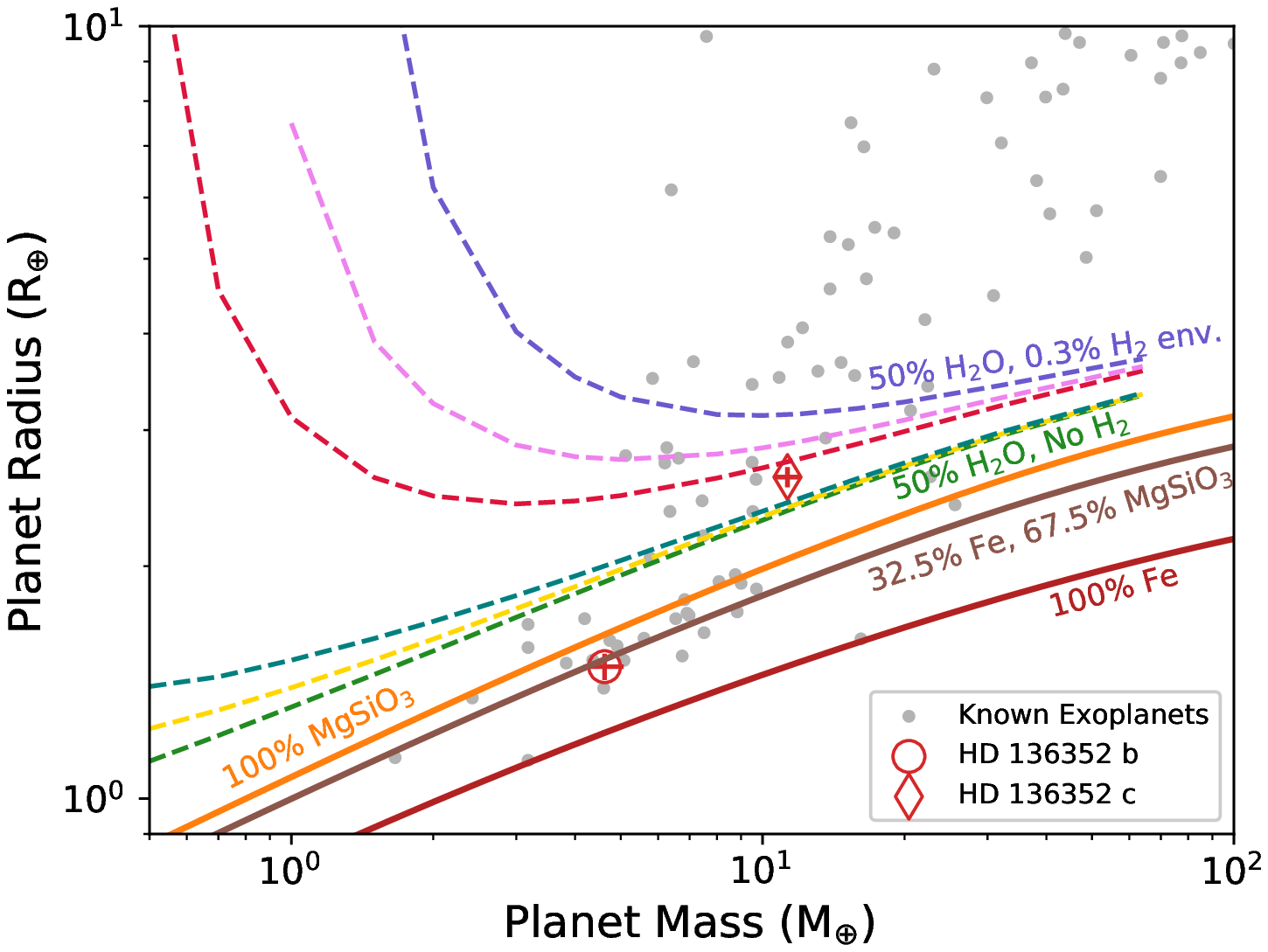} &
    \includegraphics[width=\columnwidth]{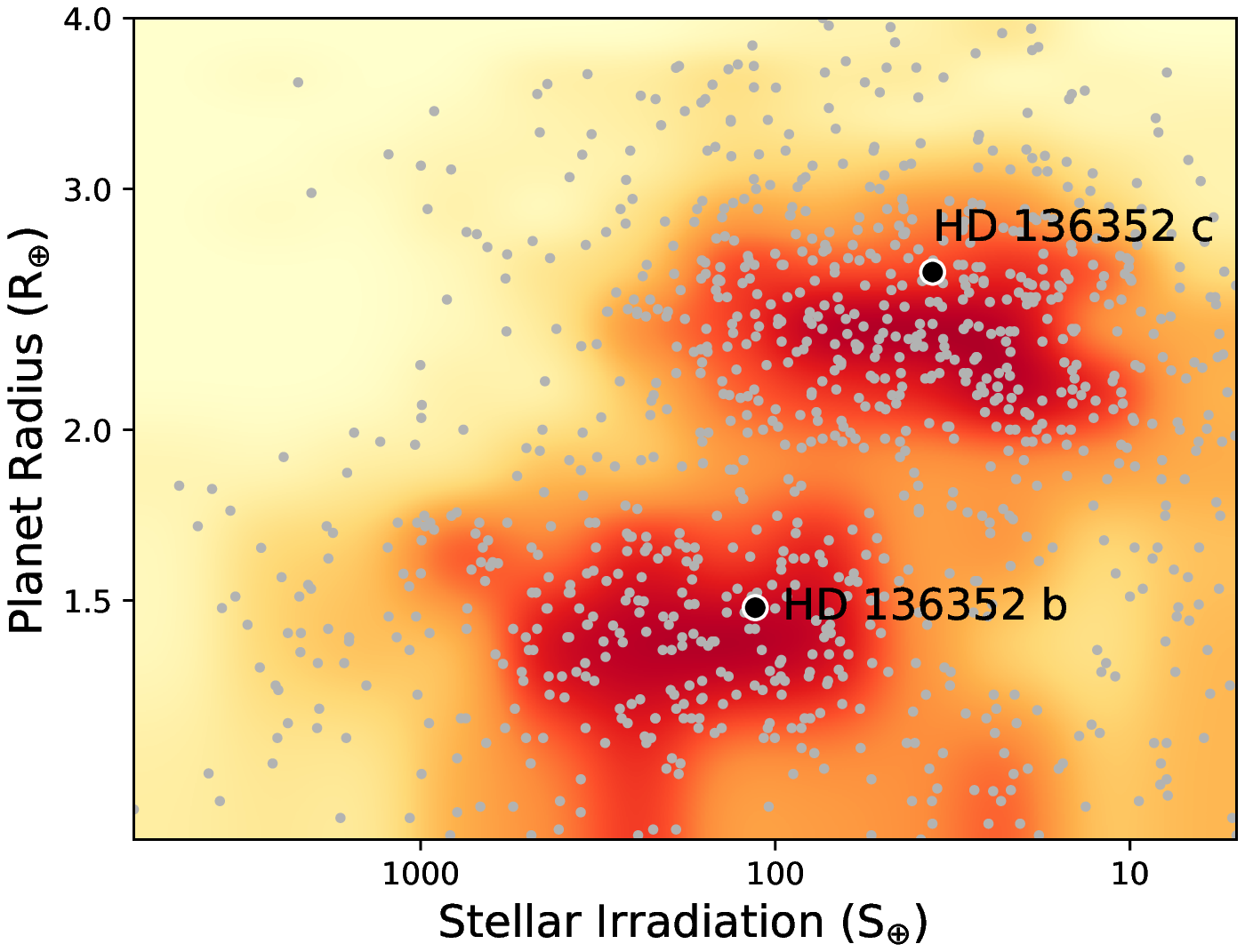}
    \end{tabular}
    \caption{Location of the HD~136352 b and c planets with respect to
      exoplanet demographic populations. {\it Left:} Mass-radius
      diagram including known planets with mass and radius
      uncertainties less than 15\%. The locations of the b and c
      planets are marked as indicated. Also shown are models for
      planets of various compositions with or without H$_2$ envelopes
      \citep{zeng2019}. {\it Right:} Planet radius and stellar
      irradiation for a sample of well-characterized confirmed
      exoplanets (gray points) with better than 15\% precision in the
      measured radius value. The background colors describe the
      density of data points, with darker tones having more
      points. The inner transiting planets orbiting HD~136352 exist on
      opposite sides of the small planet radius gap.}
    \label{fig:mr}
\end{figure*}

In Figure~\ref{fig:mr} (left panel), we show a mass-radius diagram
that contains known exoplanets with measured masses (i.e., not $M
\sin{i}$) and radii with uncertainties less than 15\%. Figure
\ref{fig:mr} also includes modeled planetary composition models for
rocky planets with or without H$_2$ envelopes from
\citet{zeng2019}. Our data analyses from Section~\ref{exofast} show
that the inner planet of HD~136352 is relatively small, with a radius
of 1.482~$R_\oplus$. This, when compared with the mass of
4.62~$M_\oplus$ yields a bulk density of
$7.8^{+1.2}_{-1.1}$~gcm$^{-3}$ (see Table~\ref{tab:planet}). A
comparison to the 5.5~gcm$^{-3}$ bulk density of Earth suggests that
planet b has a dense core that is potentially iron dominated. Indeed,
the planet b density lies near the peak density for rocky planets
based on the empirical predictions of \citet{weiss2014}. Planet c, on
the other hand, has a significantly larger radius of 2.608~$R_\oplus$,
yielding a density of only $3.50^{+0.41}_{-0.36}$~gcm$^{-3}$ which is
consistent with a thick hydrogen-helium envelope. This can be
explained through the photo-evaporation hypothesis that produces the
radius valley in the exoplanet distribution
\citep{lopez2013,owen2013a,fulton2017}. According to this hypothesis,
planet b would consist of a ``bare" core, which, due to the proximity
to the host star, would have suffered atmospheric stripping within
100~Myr of formation. HD~136352c, on the other hand, received less
high-energy radiation and maintained a thick atmospheric
envelope. Given the mass and radius for both planets, there are
significant constraints for inferences to their interior structures in
the context of composition. Namely, the stellar elemental abundances
(see Section~\ref{system}) would need to be measured to a higher
precision, 0.02--0.04~dex \citep{hinkel2018}, necessary to
meaningfully constrain the mineralogy \citep{unterborn2019}. In
addition, fundamental to all of the mass-radius models are critical
assumptions regarding the composition of rocky exoplanets and the
underlying mineral physics. These assumptions typically cause over- or
under-predictions in empirical models \citep[e.g.,][]{zeng2016a} when
characterizing ultra-high pressures present in the cores of
super-Earths and mini-Neptunes \citep{unterborn2019}.

To explore the evaporation hypothesis further, we produced the
radius-insolation diagram shown in Figure~\ref{fig:mr} (right panel),
including the planet radius versus stellar irradiation relative to the
Earth ($S_{\earth}$) for a sample of well characterized confirmed
exoplanets. Starting from all confirmed exoplanets listed on the NASA
Exoplanet
Archive\footnote{\url{https://exoplanetarchive.ipac.caltech.edu/}}, we
excluded those with a controversial flag, those with only a limit for
planet radius, and those with planetary radii with errors greater than
15\%. For the remaining planets, we either used the value of
$S_{\earth}$ provided or we calculated this value using the available
stellar and orbital properties. This well-characterized sample cleanly
displays a gap in the planetary radius distribution
\citep[e.g.,][]{fulton2017}. Interestingly, the inner transiting
planets orbiting HD~136352 straddle this gap despite differing in
stellar irradiation by only a factor of three (at the present
time). This bifurcation in planet properties makes the HD~136352
system an excellent laboratory for testing the cause of the radius gap
for small planets. Indeed, the two planets tend to have approximately
the planetary radii suggested by the aforementioned evaporation model
by \citep{owen2013a}: $1.3R_\oplus$ for a "stripped" core and
$2.6R_\oplus$ for a mini-Neptune.

With the host star being similar to the Sun, the known HD~136352
planets lie far interior to the inner boundaries of the Habitable Zone
\citep{kasting1993a,kopparapu2013a,kopparapu2014,kane2016c}, but do
lie within the Venus Zone \citep{kane2014e}. This is mostly relevant
to planet b, and other terrestrial planets that may be present within
the system, because the exploration of planetary habitability and
comparative planetology aims to study the major factors that drive the
bifurcation of habitable versus uninhabitable environments
\citep{hamano2013,kane2019d,way2020}. Terrestrial planets orbiting
close to a bright host star, such as those discussed here, provide the
best opportunities to conduct the needed atmospheric studies to inform
the diversification processes \citep{ostberg2019}.


\section{Atmospheric Characterization}
\label{atmos}

In order to quantify the expected follow-up potential to observe the
atmospheres of HD~136352 b and c, we calculated their estimated
transmission spectroscopy signal-to-noise ratios (S/N) using the
transmission spectroscopy metric (TSM) developed by
\citet{kempton2018}. This metric is dependent on the planet radius,
mass, and equilibrium temperature, as well as the stellar radius and
apparent $J$-band magnitude. The TSM method also includes a scale
factor that is dependent on the radius of the planet that allows the
TSM values to have a 1:1 ratio with simulated Near Infrared Slitless
Spectrograph (NIRISS) results produced by \citet{louie2018}, which
assumes 10 hours of observations with NIRISS on board the James Webb
Space Telescope ({\it JWST}). By applying the values for both of the
HD~136352 transiting planets obtained through Tables~\ref{tab:star}
and \ref{tab:planet}, we find that the estimated TSM values for the b
and c planets are 12 and 148, respectively.

To provide context for the estimated atmospheric observability of
HD~136352b and c, we compared their TSM values to those of the
TRAPPIST-1 planets. Using the stellar and planetary parameters from
\citet{gillon2017a}, we calculated the TSM values for TRAPPIST-1b--g
to be 44, 21, 24, 23, 27, and 15, respectively. This illustrates that
HD~136352b would be expected to require more observation time to
achieve the same S/N as the TRAPPIST-1 planets, while HD~136352c would
require far less time. The stark difference in estimated S/N between
the HD~136352 planets is due to their differences in planetary radii
and the expected compositions of their atmospheres. Since the TSM
calculation is proportional to the planetary radius to the third
power, the larger radius of HD~136352c gives it a steep increase in
estimated S/N. Furthermore, HD~136352b is not expected to have a
hydrogen-dominated atmosphere, and thus the original simulations of
\citet{louie2018} assume a mean molecular weight that is nearly ten
times larger for this type of planet than for planets like HD~136352c
($\mu$\,=\,18 vs. 2.3). This leads to a correspondingly smaller S/N
due to the inverse linear dependence of transmission spectrum feature
sizes with atmospheric mean molecular weight
\citep{millerricci2009a}. Even so, prospects exist for potential
detection of an extended atmosphere for HD~136352c, through mechanisms
such as Helium absorption \citep{allart2018}.

It should be noted that HD~136352's $K$-band magnitude of 4.159 may
result in observations by {\it JWST} being difficult because the
saturation limit for spectroscopy is $K \sim 4$
\citep{beichman2014b}. Therefore, ground-based high resolution
transmission spectroscopy using the cross correlation method
\citep{snellen2010b} may be a more productive avenue to pursue because
this technique is ideally suited to planets orbiting very bright host
stars. The TSM values we calculated should also be proportional to the
S/N expected in the near-infrared with this technique.


\section{Conclusions}
\label{conclusions}

RV exoplanet systems are among the best-characterized systems in the
overall exoplanet inventory. This is because the brightness of the
host stars enables significant observational capability and ancillary
science, including the study of planetary orbits, architectures, and
interactions \citep{ford2014,kane2014b,winn2015}. The ancillary
science includes investigations of the radius gap, evaporation
scenarios, and the structure of planets that span the planetary radius
gap \citep{mousis2020a}. Furthermore, the relative proximity of the RV
systems makes them attractive targets for direct imaging surveys that
aim to directly detect the known planets
\citep{kane2013c,kane2018c,kopparapu2018}. Thus, when planets in RV
systems are also found to transit their host star, they become truly
exceptional in the scope of possible science, particularly when
multiple planets are found to transit in the same system.

The trajectory of exoplanetary science is leading toward the
characterization of planetary atmospheres. In order to fully exploit
the potential of transmission spectroscopy techniques, numerous
excellent targets are required that orbit bright host stars. The
HD~136352 system is now known to harbor two transiting planets, and
our analysis has determined that the radii of these planets place them
on either side of the well-known radius gap. As described in
Section~\ref{atmos}, the imperative to understand the atmospheric
evolution of such planets makes them attractive follow-up targets for
atmospheric studies. We have shown that planet c is an especially
promising target in terms of the expected S/N from both transmission
and emission spectroscopy observations that could be carried out with
{\it JWST}.

One pressing concern is that, despite long-term constraints from RVs
and from the single transit, the period of planet c is still
relatively uncertain. Therefore, further observations, either during
the {\it TESS} extended mission or by the {\it CHEOPS} mission
\citep{broeg2014}, are needed to ensure that the ephemeris of this
planet can be refined to enable followed-up observations
\citep{dragomir2020}. As noted in Section~\ref{exofast}, planet d did
not pass through inferior conjunction during the {\it TESS}
observations, and so follow-up photometric campaigns could reveal
whether planet d also transits.

As described in Section~\ref{intro}, the transit detection of known RV
planets has historically provided some of the most interesting
exoplanets over the past two decades. This work demonstrates that this
is still true, and the advantage of RV observations has enabled us to
provide significant mass constraints, due to the legacy of RV
observations that preceded the transit detections. With the {\it TESS}
mission transitioning into a mode whereby it returns to previously
observed parts of the sky, we can expect that there will be further
opportunities to uncover new insights into the known exoplanetary
systems.


\section*{Acknowledgements}

The authors would like to thank Jason Eastman for helpful instruction
regarding the use of {\sc EXOFASTv2}, and the anonymous referee whose feedback improved the manuscript. S.R.K. acknowledges support by
the National Aeronautics and Space Administration through the TESS
Guest Investigator Program (17-TESS17C-1-0004). P.D. acknowledges
support from a National Science Foundation Astronomy and Astrophysics
Postdoctoral Fellowship under award AST-1903811. S.U, F.B., X.D., D.E,
C.L., F.P., M.M., and D.S. acknowledge the financial support of the
National Center for Competence in Research, PlanetS, of the Swiss
National Science Foundation (SNSF). D.E. acknowledges support from
from the European Research Council (ERC) under the European Union’s
Horizon 2020 research and innovation programme (project {\sc Four
  Aces}; grant agreement No 724427). T.L.C. acknowledges support from
the European Union's Horizon 2020 research and innovation programme
under the Marie Sk\l{}odowska-Curie grant agreement No.~792848
(PULSATION). H.R.A.J. is supported by UK STFC grant ST/R006598/1. We
acknowledge the traditional owners of the land on which the AAT
stands, the Gamilaraay people, and pay our respects to elders past and
present. We gratefully acknowledge the efforts and dedication of the
Keck Observatory staff for support of HIRES and remote observing. We
recognize and acknowledge the cultural role and reverence that the
summit of Maunakea has within the indigenous Hawaiian community. We
are deeply grateful to have the opportunity to conduct observations
from this mountain.  Funding for the {\it TESS} mission is provided by
NASA's Science Mission directorate. This research has made use of the
Exoplanet Follow-up Observation Program website, which is operated by
the California Institute of Technology, under contract with the
National Aeronautics and Space Administration under the Exoplanet
Exploration Program. Resources supporting this work were provided by
the NASA High-End Computing (HEC) Program through the NASA Advanced
Supercomputing (NAS) Division at Ames Research Center for the
production of the SPOC data products. This paper includes data
collected by the TESS mission, which are publicly available from the
Mikulski Archive for Space Telescopes (MAST). The research shown here
acknowledges use of the Hypatia Catalog Database, an online
compilation of stellar abundance data as described in
\citet{hinkel2014}, which was supported by NASA's Nexus for Exoplanet
System Science (NExSS) research coordination network and the
Vanderbilt Initiative in Data-Intensive Astrophysics (VIDA). This
research has made use of the NASA Exoplanet Archive, which is operated
by the California Institute of Technology, under contract with the
National Aeronautics and Space Administration under the Exoplanet
Exploration Program. The results reported herein benefited from
collaborations and/or information exchange within NASA's Nexus for
Exoplanet System Science (NExSS) research coordination network
sponsored by NASA's Science Mission Directorate.


\software{EXOFAST \citep{eastman2013,eastman2020}}




\end{document}